\documentclass[
  aps,prl,twocolumn,
  amsmath,amssymb,
  superscriptaddress,
  nofootinbib
]{revtex4-2}

\usepackage{graphicx}
\usepackage{bm}
\usepackage{hyperref}
\makeatletter
\let\frontmatter@footnote@produce@endnote\frontmatter@footnote@produce@footnote
\makeatother

\begin{document}

\title{Geometric Search for Hawking Radiation from Nearby Primordial Black Holes}

\author{Shuo Xiao}
\affiliation{School of Physics and Electronic Science, Guizhou Normal University, Guiyang 550001, People’s Republic of China}

\author{Shuang-Nan Zhang*}
\affiliation{Key Laboratory of Particle Astrophysics, Institute of High Energy Physics, Chinese Academy of Sciences, Beijing 100049, China}
\affiliation{University of Chinese Academy of Sciences, Chinese Academy of Sciences, Beijing 100049, China}

\date{\today}

\begin{abstract}
A nearby primordial-black-hole (PBH) evaporation burst would produce a curved gamma-ray wavefront, leading to detectable departures from plane-wave inter-satellite delays. We introduce a purely geometric method that combines imaging localizations with multi-spacecraft timing to determine the distance of a gamma-ray transient. Applied to \textit{Swift}-localized short GRBs, the current sample shows no significant deviation from the plane-wave expectation, with the most constraining event reaching $1.2$~AU and already probing a meaningful Solar-System-scale regime. Our analysis shows that direct distance measurements are achievable to $10^3$~AU scales with the current and near-future technical capabilities. 
Once a finite source distance is measured, the corresponding PBH mass and lifetime can be directly inferred. Future wide-field localization and long-baseline deep-space gamma-ray detectors could extend such searches to $10^5$~AU and beyond.
\end{abstract}

\maketitle
{\let\thefootnote\relax\footnotetext{* Corresponding author: zhangsn@ihep.ac.cn}}

\paragraph{Introduction.—}
High-energy transient timing analyses almost universally adopt the plane-wave approximation, because the vast majority of gamma-ray bursts (GRBs) are established to be cosmological through afterglows, host associations, and in a few cases multi-messenger counterparts~\cite{Costa1997,vanParadijs1997,Piran2004,woosley2006supernova,Berger2014,Kumar2015,abbott2017gravitational,goldstein2017ordinary,savchenko2017integral,li2018insight}. This approximation underpins multi-spacecraft triangulation and Interplanetary Network (IPN) localizations~\cite{1999ApJS..122..497H,McClanahan2011,Svinkin2022}. Nevertheless, several speculative scenarios could produce nearby, short-duration gamma-ray flashes, among which the final evaporation of primordial black holes (PBHs) is a particularly well-motivated example~\cite{Hawking1974,Page1976,Carr2010,MacGibbon1991,Halzen1991,ukwatta2016investigation}. A nearby burst would imprint a distinctive geometric signature: the incident wavefront is curved, so arrival-time differences across widely separated spacecraft depart from plane-wave expectations.

Timing-only reconstructions suffer from a fundamental degeneracy, because changes in source direction can mimic changes in distance~\citep{ukwatta2016investigation}. We overcome this limitation by combining imaging localizations with inter-satellite timing. Once the sky direction is fixed, the finite-distance problem reduces to a single geometric parameter, the source distance. We then formulate, for the first time, a quantitative framework for direct event-by-event distance inference from this combination and apply it systematically to localized short GRBs in current multi-mission data.

This is particularly important for PBH searches, because the key question is whether a burst lies at a finite, non-cosmological distance. Such a regime is astrophysically well motivated: within the Solar System and out to the nearest stellar distances, few known sources other than exotic nearby transients are expected to produce bright millisecond-to-second gamma-ray flashes. Ordinary stellar flares are much slower, while planetary high-energy phenomena such as terrestrial gamma-ray flashes~\citep{Fishman1994} are far too faint to mimic a bright GRB-like event at AU scales. A burst established at such nearby distances would therefore strongly favor a PBH interpretation and, once its distance is measured, would enable direct event-by-event physical inference. 

\begin{figure*}
\centering
\includegraphics[width=0.85\linewidth]{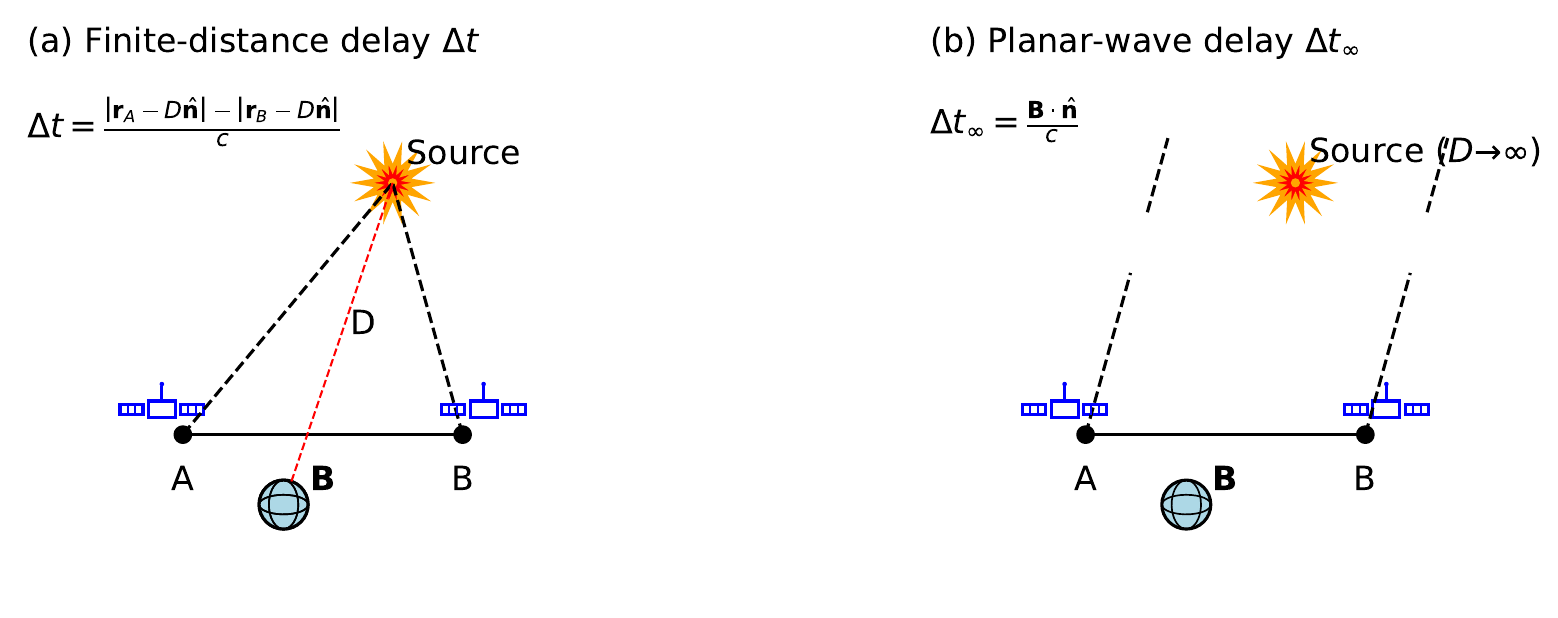}
\caption{Geometric test for finite-distance gamma-ray transients. Imaging localization fixes $\hat{\mathbf{n}}$. For a finite distance $D$, the delay follows Eq.~(\ref{eq:finite}); in the plane-wave limit it reduces to Eq.~(\ref{eq:plane}). Curvature induces a small deviation scaling as $|\delta t|\propto B_\perp^2/D$, enabling direct geometric distance measurements once the curvature signal is resolved.}
\label{fig:schem}
\end{figure*}

\begin{figure*}
\centering
\includegraphics[width=0.7\linewidth]{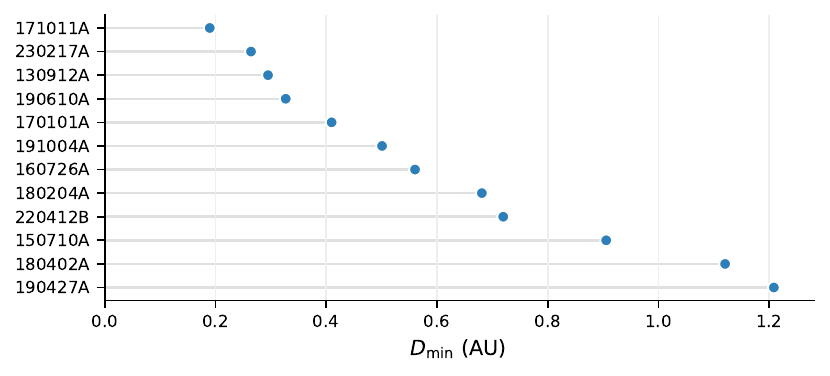}
\caption{Distance constraints from the non-detection of wavefront curvature for short GRBs jointly observed by \textit{Konus}-Wind and BAT. Each point shows the conservative $3\sigma$ lower limit $D_{\min}$ inferred for an individual event. The most constraining case reaches $D_{\min}\approx1.2$~AU, showing that current observations have already entered the Solar-System-scale regime relevant to nearby PBH searches. Although the present sample mainly yields lower limits, it already reaches the distance scale at which direct geometric searches for nearby PBH bursts become meaningful.}
\label{fig:Dmin_wind}
\end{figure*}

\begin{figure*}
\centering
\includegraphics[width=0.7\linewidth]{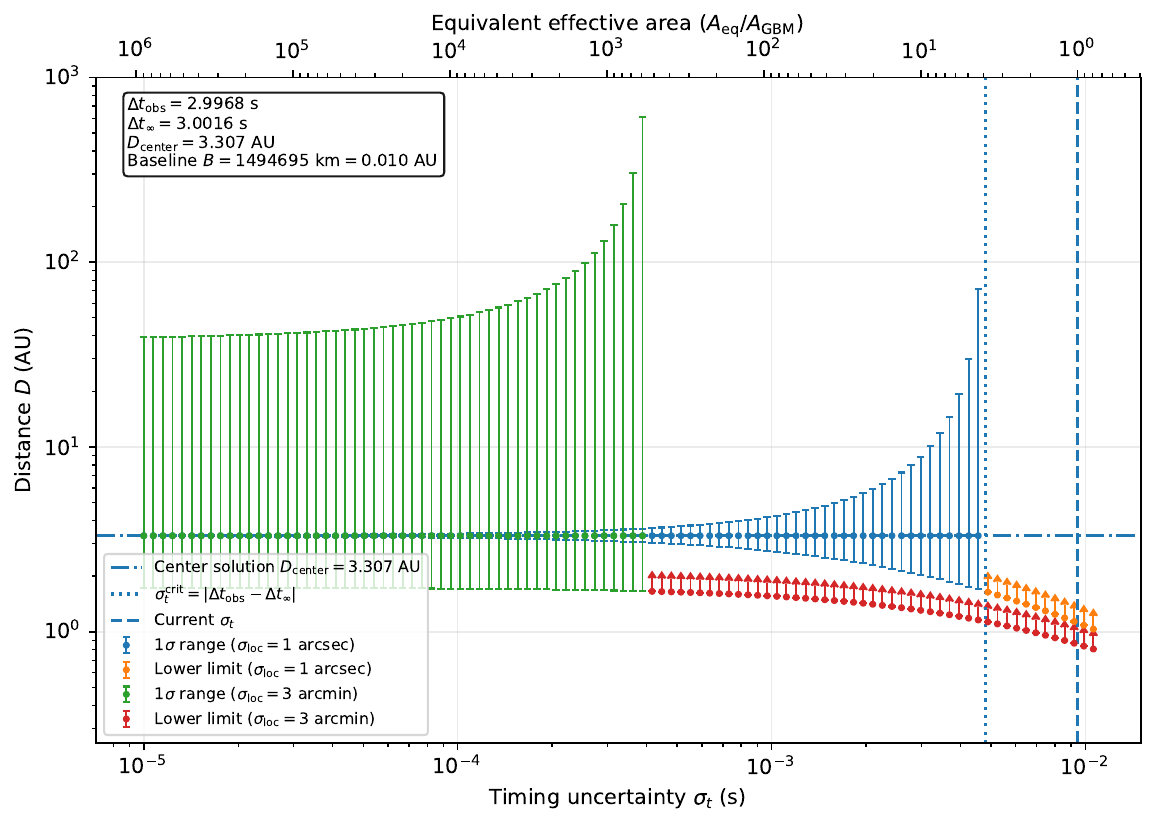}
\caption{Distance constraint versus timing uncertainty $\sigma_t$ for GRB~160726A. Blue points show the central solution and $1\sigma$ interval for $\sigma_{\rm loc}=1$ arcsec, while orange and green points show the corresponding $1\sigma$ lower limits for $\sigma_{\rm loc}=1$ arcsec and $3$ arcmin once the upper bound becomes unbounded. The horizontal dash-dotted line marks $D_{\rm center}=3.307$~AU. The vertical dotted and dashed lines mark $\sigma_t^{\rm crit}=|\Delta t_{\rm obs}-\Delta t_\infty|$ and the current timing uncertainty, respectively. The top axis shows the equivalent effective area relative to GBM. This figure illustrates the threshold at which the method changes from yielding only consistency with the infinite-distance limit to directly measuring a finite source distance.}
\label{fig:single_event}
\end{figure*}

\begin{figure*}
\centering
\includegraphics[width=0.7\linewidth]{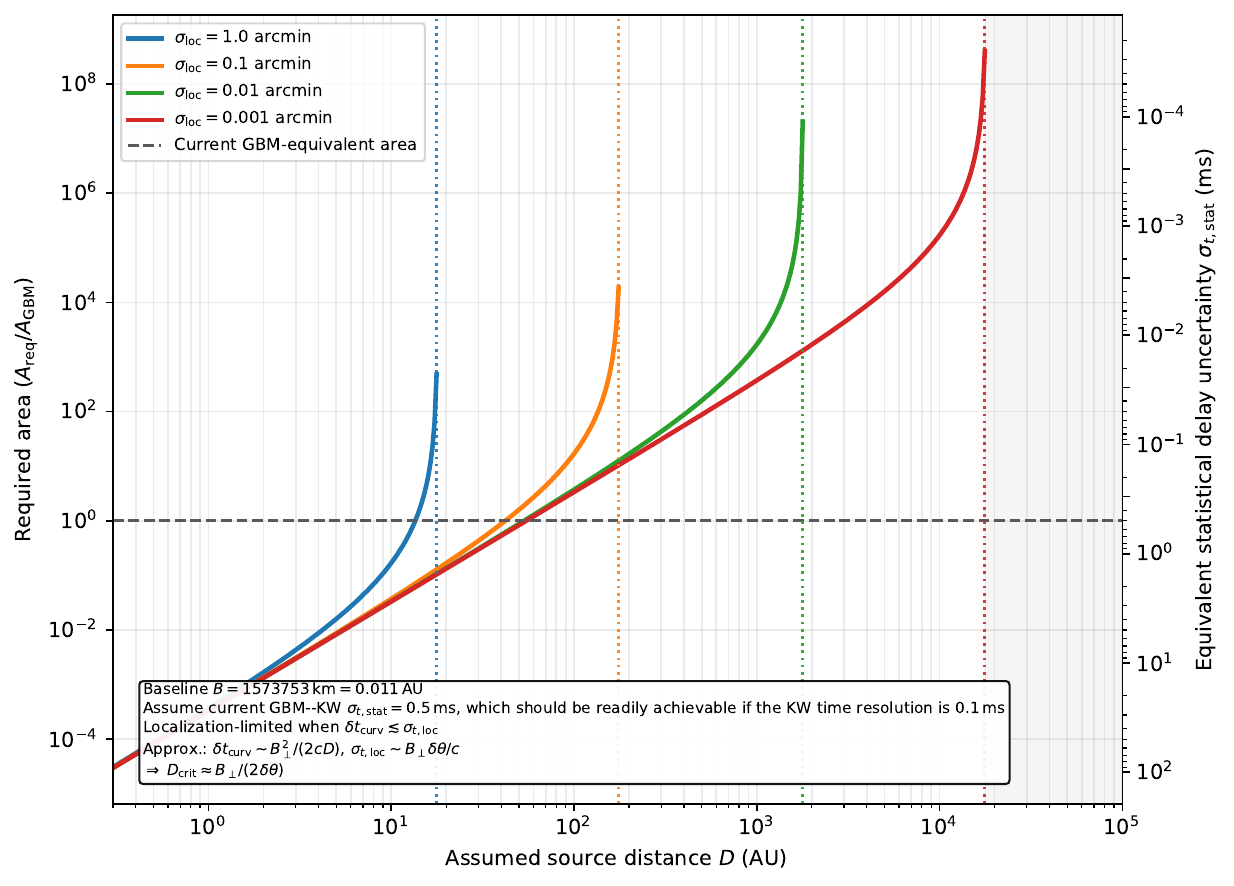}
\caption{Required effective area versus assumed source distance for different localization uncertainties. The curves show the equivalent area, in units of the current GBM effective area, required to retain a finite upper bound on the source distance for the GBM--KW baseline. We adopt $\sigma_{t,\rm stat}=0.5$~ms, consistent with short-GRB timing at $\sim$0.1~ms time resolution~\cite{xiao2021,xiao2022ground}. The horizontal dashed line marks the current GBM-equivalent area, and the vertical dotted lines mark the critical distances $D_{\rm crit}$ at which the constraint becomes localization limited. The right axis gives the equivalent statistical delay uncertainty. This figure quantifies how improved timing sensitivity and especially better localization expand the distance range over which the method can directly measure, rather than merely constrain, the source distance.}
\label{fig:reqmap}
\end{figure*}

\paragraph{Geometric timing model.—}
Figure~\ref{fig:schem} shows the geometric setup. For two detectors $i$ and $j$ at barycentric positions $\mathbf r_i$ and $\mathbf r_j$, the plane-wave delay for a source from direction $\hat{\mathbf n}$ is
\begin{equation}
\Delta t_{ij}^{\infty}=\frac{\mathbf B_{ij}\!\cdot\!\hat{\mathbf n}}{c},\qquad
\mathbf B_{ij}\equiv \mathbf r_i-\mathbf r_j ,
\label{eq:plane}
\end{equation}
where $c$ is the speed of light. For a source at a finite distance $D$ along $\hat{\mathbf n}$, the delay becomes
\begin{equation}
\Delta t_{ij}(D)=\frac{\left|\mathbf r_i-D\hat{\mathbf n}\right|-\left|\mathbf r_j-D\hat{\mathbf n}\right|}{c}.
\label{eq:finite}
\end{equation}
Once an imaging localization fixes $\hat{\mathbf n}$, the problem reduces to a single parameter, $D$. Geometrically, a fixed delay defines a two-sheeted hyperboloid with the detectors as foci, while the localization fixes a ray along $\hat{\mathbf n}$; the finite-distance solution is their intersection.

For a measured delay $\Delta t_{ij}^{\rm obs}$, Eq.~(\ref{eq:finite}) can be inverted analytically. Defining
\begin{equation}
\Delta_{ij}\equiv c\,\Delta t_{ij}^{\rm obs},\qquad
\alpha_i\equiv \hat{\mathbf n}\cdot \mathbf r_i,\qquad
\alpha_j\equiv \hat{\mathbf n}\cdot \mathbf r_j,
\end{equation}
the corresponding source distance is
\begin{equation}
D=
\frac{
\left(|\mathbf r_i|^2-|\mathbf r_j|^2\right)(\alpha_i-\alpha_j)
-\Delta_{ij}^2(\alpha_i+\alpha_j)
\pm
\Delta_{ij}\sqrt{\mathcal{Q}_{ij}}
}{
2\left[(\alpha_i-\alpha_j)^2-\Delta_{ij}^2\right]
},
\label{eq:D_exact}
\end{equation}
with
\begin{align}
\mathcal{Q}_{ij}={}&
|\mathbf r_i|^4+|\mathbf r_j|^4
-2|\mathbf r_i|^2|\mathbf r_j|^2
-2\Delta_{ij}^2\left(|\mathbf r_i|^2+|\mathbf r_j|^2\right)
\nonumber\\
&
+\Delta_{ij}^4
-4|\mathbf r_i|^2\alpha_i\alpha_j
+4|\mathbf r_i|^2\alpha_j^2
+4|\mathbf r_j|^2\alpha_i^2
\nonumber\\
&-4|\mathbf r_j|^2\alpha_i\alpha_j
+4\alpha_i\alpha_j\,\Delta_{ij}^2 .
\label{eq:Q_exact}
\end{align}
The physical solution is the branch that satisfies Eq.~(\ref{eq:finite}) with $D>0$.

Observed delays $\Delta t_{ij}^{\rm obs}$ are measured with the modified cross-correlation function (MCCF)  \cite{li1999temporal,li2001timing,li2004timescale,xiao2021}, and uncertainties are estimated from Monte Carlo light-curve realizations. Defining the plane-wave residual $\delta t_{ij}\equiv \Delta t_{ij}^{\rm obs}-\Delta t_{ij}^{\infty}$ and $B_{\perp}\equiv|\mathbf{B}_{ij}\times\hat{\mathbf{n}}|$, we combine the MCCF uncertainty $\sigma_{ij}$ with localization propagation as
\begin{equation}
\sigma_{ij,{\rm tot}}^2=\sigma_{ij}^2+\sigma_{ij,{\rm loc}}^2,\qquad
\sigma_{ij,{\rm loc}}\simeq \frac{B_{\perp}}{c}\,\sigma_\theta,
\label{eq:sigtot}
\end{equation}
where $\sigma_\theta$ is the conservatively isotropic localization uncertainty. 
We then quantify consistency with the plane-wave hypothesis through $x_{ij}\equiv \delta t_{ij}/\sigma_{ij,{\rm tot}}$.

For $D\gg |\mathbf{B}_{ij}|$, Eq.~(\ref{eq:finite}) yields a curvature correction of order $B_\perp^2/(cD)$, so that the finite-distance deviation scales as $|\delta t|\propto B_\perp^2/D$. Using the conservative bound $|\delta t|\lesssim B_\perp^2/(2cD)$, a null detection at the $k\sigma$ level implies
\begin{equation}
D_{\min}\simeq \frac{B_\perp^{2}}{2c\,k\,\sigma_{ij,{\rm tot}}},\qquad (k=3).
\label{eq:Dmin}
\end{equation}
What matters physically, however, is not the lower limit itself but whether the event remains consistent with the infinite-distance limit or instead admits a finite distance scale. Once the curvature signal is resolved above the combined timing and localization uncertainties, Eq.~(\ref{eq:finite}) directly yields the source distance and thus distinguishes a nearby burst from an ordinary cosmological GRB.

\paragraph{Sample and analysis choices.—}
We select short GRBs ($T_{90}<3$ s) localized by imaging instruments, focusing on \textit{Swift}/BAT~\cite{sakamoto2008first}. To emphasize cases not independently established as cosmological, we restrict the sample to bursts without measured redshift or secure host association. Our primary demonstration set comes from events jointly observed by \textit{Swift}/BAT and \textit{Konus}-Wind~\cite{aptekar1995konus}, leveraging the long Wind--Earth baseline for maximal curvature sensitivity within the currently available sample. The final BAT--KW sample contains 12 short GRBs.

We measure cross-instrument delays in matched bands and restrict the MCCF to the prompt-emission interval. As internal consistency checks in the near-Earth network, we also examine pairs such as GBM--BAT, GBM--HXMT/HE, and GBM--GECAM when available~\cite{Meegan2009,zhang2020overview,xiong2020gecam}.

\paragraph{Results: plane-wave validation and feasibility of finite-distance identification.—}
For the 12 BAT--KW events in our main sample, all measured delays are consistent with the plane-wave prediction within $3\sigma$ once timing and localization uncertainties are included, showing no significant evidence for wavefront curvature. Near-Earth cross-checks using GBM--BAT, GBM--HXMT/HE, and GBM--GECAM, when available, are likewise consistent with the plane-wave prediction within $3\sigma$.

In the present sample, the combination of source distances, detector baselines, timing uncertainties, and localization precision means that most events do not yield resolved finite-distance solutions. Even so, the most constraining event reaches $D_{\min}\approx1.2$~AU (Figure~\ref{fig:Dmin_wind}), showing that current observations have already entered the Solar-System-scale regime in which a genuinely finite nearby distance could in principle be identified. This lower bound is not itself the main PBH discriminator; rather, it shows that the method has reached the distance scale where a finite, non-cosmological burst distance would become physically meaningful. The current non-detections therefore reflect the limitations of the available sample and detector configurations rather than that of the method itself. Our analysis show that finite nearby distances are already measurable in principle from AU to $10^3$~AU scales with current capabilities.

\paragraph{Roadmap to direct geometric distance measurements out to $10^{5}$~AU and beyond.—}
The same geometry provides a practical path to extend direct PBH-inspired distance measurements far beyond AU scales. For nearby events, the main challenge is timing, but at larger distances the measurement increasingly becomes localization limited. This transition is illustrated by Figures~\ref{fig:single_event} and \ref{fig:reqmap}, and can be characterized by
\begin{equation}
D_{\rm crit}\approx \frac{B_\perp}{2\delta\theta},
\label{eq:Dcrit}
\end{equation}
where $B_\perp$ is the projected baseline and $\delta\theta$ is the localization uncertainty. Direct distance measurements require the curvature signal to be resolved before this limit is reached. For PBH searches, the physically decisive outcome is therefore not a lower limit by itself, but the identification of a finite, non-cosmological distance scale.

This framework immediately translates into a staged observational roadmap. For a Wind--Earth baseline of order $B_\perp\sim10^6$~km, sub-ms matched-band timing together with arcminute localization, building on the wide-field MPO capability already demonstrated by Einstein Probe/WXT~\cite{Yuan2022}, already gives access to distances of order $10$ AU, i.e. deep into the Solar-System regime. This is essentially achievable with present-day technology: the main limitation is not the baseline itself, but the lack of sufficiently high-time-resolution public products from \textit{Konus}-Wind. Improving the localization by one order of magnitude, to $\sim0.1$~arcmin as expected for a next-generation MPO, would push the same baseline toward the $10^2$~AU regime, beyond which the measurement would increasingly become localization limited, as indicated by Figure~\ref{fig:reqmap}. Even without such an improvement, extending the baseline to Sun--Earth scales while retaining arcminute localization would already push the directly measurable distance range to of order $10^3$~AU.

Longer baselines then open a much larger search volume. For $B_\perp$ of order 1~AU, a Mars--Earth configuration combined with $\sim0.1$~arcmin localization could extend the directly measurable distance range to $10^4$~AU, making Mars missions especially attractive opportunities for piggyback gamma-ray detectors. An even more powerful configuration would use Sun--Earth-scale baselines together with a new generation of wide-field instruments reaching arcsecond localization; in that case, the localization-limited reach can approach or exceed $10^5$~AU. Since both Mars and solar missions are likely to continue internationally, they offer realistic opportunities to push purely geometric PBH searches far beyond the current AU-scale frontier.

In practice, $\sigma_t$ is statistics driven for background-dominated short transients, with $\sigma_t\propto{\rm SNR}^{-1}\propto A_{\rm eff}^{-1/2}$ for comparable backgrounds and bandpasses. This motivates both larger effective area and lower background on millisecond timescales. Absolute timing is already adequate for this program: near-Earth missions provide $\mu$s--$100~\mu$s absolute timing accuracy and $\mu$s--sub-$\mu$s event time tagging~\cite{Meegan2009,liu2020high,xiao2022ground}, well below the statistical delay uncertainties relevant here. For \textit{Konus}-Wind, the main limitation is typically the publicly available light-curve binning rather than clock accuracy~\cite{aptekar1995konus}. Thus, the principal technological drivers are no longer clock performance, but wide-field localization accuracy, high-time-resolution data products, and access to long baselines.

A central advantage of the geometric approach is that, once a finite source distance $D$ is established for a candidate event, the observed energy flux $F$ immediately gives the intrinsic luminosity, $L=4\pi D^2F$. Under the interpretation that the burst is Hawking radiation from a nonrotating PBH, and defining $F_{-6}\equiv F/(10^{-6}\ {\rm erg\ cm^{-2}\ s^{-1}})$ (the typical peak flux of a short GRB), the corresponding PBH mass is
\begin{equation}
M\simeq 1.54\times10^{-12}
\left(\frac{D}{100\,{\rm AU}}\right)
F_{-6}^{-1/2}
M_{\rm Moon},
\label{eq:HawkingMass}
\end{equation}
where $M_{\rm Moon}$ is the lunar mass. The corresponding PBH lifetime from formation to final evaporation is then
\begin{equation}
\tau\simeq 3.80\times10^{9}
\left(\frac{D}{100\,{\rm AU}}\right)^3
F_{-6}^{-3/2}
{\rm yr}.
\label{eq:HawkingLifetime}
\end{equation}
Thus, unlike rate-based searches, a geometric distance measurement would provide direct event-by-event access to the PBH mass and lifetime under the Hawking-burst interpretation~\cite{Hawking1974,Hawking1975}. It is interesting to note that for $\tau\sim1.37\times 10^{10}$ yr, i.e., the age of the universe, we have $D\sim 150$ AU, within the reach of an AU scale baseline with current wide-field X-ray imaging and gamma-ray timing capabilities.

\paragraph{Summary.—}
We have introduced a purely geometric method for measuring the distance of nearby gamma-ray transients by combining imaging localizations with multi-spacecraft timing. Applied to localized short GRBs without redshift, the current sample shows no evidence for wavefront curvature and remains consistent with the plane-wave approximation across missions. This reflects the limitations of the presently available events and detector configurations, rather than a limitation of the method itself.

The main result is that AU-scale direct distance measurements are already feasible in principle with current or near-future capabilities. What matters physically is not the lower limit by itself, but whether the event can be shown to lie at a finite, non-cosmological distance. Such a result would immediately distinguish the burst from ordinary cosmological GRBs and would strongly point to an exotic nearby origin such as PBH evaporation. Because few known astrophysical sources are expected to produce bright millisecond-to-second gamma-ray flashes within the Solar System and its neighborhood, systematic searches for such nearby events are already well motivated.

Figures~\ref{fig:single_event} and \ref{fig:reqmap} summarize the key requirements for extending this capability. Wind--Earth-scale baselines with arcminute to sub-arcminute localization already reach from $10$ to $10^2$~AU, while Sun--Earth-scale baselines with arcminute localization can already extend the directly measurable range to $10^3$~AU. Mars--Earth configurations combined with $\sim0.1$~arcmin localization can further extend the reach to $10^4$~AU, and Sun--Earth-scale baselines combined with arcsecond localization could push direct geometric searches toward $10^5$~AU. Together, these advances provide a realistic roadmap from the current AU-scale regime to much larger search volumes for nearby PBH bursts.

\paragraph{Acknowledgments.—}
We acknowledge the public data from \textit{Fermi}/GBM, \textit{Konus}-Wind, and \textit{Swift}. This work made use of data from the \textit{Insight}-HXMT and GECAM mission, funded by the CNSA and CAS. This work is supported by the National Natural Science Foundation of China (Nos. 12303043, 12333007 and 12573043), Science and Technology Foundation of Guizhou Province (Key Program, Nos. [2025]021 and ZK[2024]430) and Space Origins Program of the Chinese Academy of Sciences.

\paragraph{Data availability.—}
The data that support the findings of this article are openly available.

\bibliography{main}

\end{document}